\documentclass[conference]{IEEEtran}
\renewcommand{\IEEEQED}{\IEEEQEDopen}

\usepackage{cite}
\ifCLASSINFOpdf
   \usepackage[pdftex]{graphicx}
   \DeclareGraphicsExtensions{.pdf}
\else
   \usepackage[dvips]{graphicx}
   \DeclareGraphicsExtensions{.eps}
\fi
\usepackage{amsmath, bm}
\newtheorem{lemma}{Lemma}
\newtheorem{claim}{Claim}
\newtheorem{definition}{Definition}
\newtheorem{theorem}{Theorem}
\newtheorem{corollary}{Corollary}
\def\E{{\rm E}}
\def\ln{{\rm ln}\,}
\newcommand{\bfp}{\bm{p}}
\newcommand{\bfe}{\bm{e}}
\newcommand{\bfd}{\bm{d}}
\usepackage{amssymb, hyperref, textgreek}
\usepackage[subnum]{cases}

\usepackage{algorithmic, algorithm}
\usepackage{fixltx2e}
\usepackage{url}
\begin{document}
\title{On Computation Rates for Arithmetic Sum}

\author{\IEEEauthorblockN{Ardhendu Tripathy and Aditya Ramamoorthy}
\IEEEauthorblockA{Department of Electrical and Computer Engineering,
Iowa State University, Ames, Iowa 50011--3060\\
Email: ardhendu@iastate.edu, adityar@iastate.edu}
}


%Telephone: (800) 555--1212, Fax: (888) 555--1212}


\maketitle

\begin{abstract}
For zero-error function computation over directed acyclic networks, existing upper and lower bounds on the computation capacity are known to be loose. In this work we consider the problem of computing the arithmetic sum over a specific directed acyclic network that is not a tree. We assume the sources to be i.i.d. Bernoulli with parameter $1/2$. Even in this simple setting, we demonstrate that upper bounding the computation rate is quite nontrivial. In particular, it requires us to consider variable length network codes and relate the upper bound to equivalently lower bounding the entropy of descriptions observed by the terminal conditioned on the function value. This lower bound is obtained by further lower bounding the entropy of a so-called \textit{clumpy distribution}. We also demonstrate an achievable scheme that uses variable length network codes and in-network compression.
%The notion of a variable-length network code is introduced and applied to a particular function computation instance over a directed acyclic network. An upper bound on its computing capacity is obtained by evaluating the entropy of a particular \textit{clumpy distribution} that fits well with the combinatorial nature of the problem. A typical set encoder that uses a variable-length network code is shown to have a better rate than that obtained by using fixed-length network codes.
\end{abstract}


% \fi
\IEEEpeerreviewmaketitle

\section{Introduction}
%This demo file is intended to serve as a ``starter file''
%for IEEE conference papers produced under \LaTeX\ using
%IEEEtran.cls version 1.8b and later.
We consider the problem of function computation \cite{giridharK05, appuFKZ11, raiD12, ramamoorthyL13} using network coding \cite{al,rm}. The setup of the problem typically is as follows. A directed acyclic graph (DAG) with capacity constraints is used to model a communication network. Certain nodes in the graph, referred to as terminals are interested in computing a target function of the data observed at some nodes (called sources) in the graph. The edges model error-free communication links and the nodes are assumed to be able to perform network coding. The objective is to find the maximum rate at which a network code can enable the terminals to compute estimates of the target function within a specified level of distortion.

%In \cite{orlitskyR01}, the authors described the optimum amount of information that needs to be transmitted for the function computation problem over two nodes. A source is connected to a terminal by a directed communication link and the terminal wants to reliably compute a function of the data locally present at the source and terminal. A block code setting was implemented, i.e., the source would encode a block of independent local data instances and transmit this information to the terminal. With the help of this information and its local data the terminal would estimate a block of target function values. Though the communication link is error-free, a vanishing block-error probability in computing the function is allowed.

The problem in its most general setting is known to be hard, and that has prompted the study of certain special cases \cite{orlitskyR01, appuFKZ11, tripathyR14, tripathyR15}. The specific case when the network has one terminal and the function needs to be computed without any distortion has received significant attention. Under this setup, one can consider either the zero-error setting or a setting where $\epsilon$-error (for arbitrary $\epsilon > 0$) error is allowed. In \cite{appuFKZ11}, the authors do not assume a joint probability distribution on the input data and focus on zero-error function computation. After describing the amount of information that needs to be transmitted for this case, the same concept is applied to cuts that separate one or more sources from the terminal in a function computation problem over a DAG. Using this approach they were able to characterize the maximum achievable computation rate and network codes that obtain it for multi-edge tree networks. The work done in \cite{kowshikK12} approaches the function computation problem in a slightly different manner. Rather than focus on a computation rate for the entire network, they look at the rate region obtained by the vector of achievable rates over each edge in the network. They consider two scenarios: \textit{worst case} and \textit{average case} complexity. The worst case complexity is related to the setting in \cite{appuFKZ11}, while the average case complexity assumes a joint probability distribution on the input data. For both scenarios, they use cut-set based arguments to characterize the rate region for tree networks. For general DAGs, cut-set based upper bounds are shown in \cite{huangTY15} to be loose. %\aditya{lower bounds?}

%We look at a problem setting that reconciles elements of \cite{orlitskyR01} with the work done in \cite{appuFKZ11}.
In this work we examine zero-error arithmetic sum computation over a specific DAG, but we assume a probability distribution on the input data (see Fig. \ref{fig:fig1}). Note that for such a network there are two distinct paths from the source $s_3$ to the terminal that allows for multitude of network coding options. In \cite{appuFKZ11}, the same network was considered in a zero-error setting, but they did not assume any distribution on the inputs. They demonstrated an upper bound on the computation rate and a matching achievable scheme. In general, the distribution of the inputs is important in defining an associated rate of computation. Indeed, if the source values are deterministic, then the actual computation does not require any information to be transmitted on the edges.

In this work, we assume that each of the source is distributed i.i.d. Bernoulli with parameter $1/2$. We demonstrate that in this setting, upper bounding the computation rate is significantly harder because of the possibility of compressing the intermediate transmissions. In addition, the presence of multiple paths for source $s_3$ allows for many ways in which the information transfer and compression can be performed. Our upper bounds on the computation rate stem from the study of the entropy of the distribution of the descriptions transmitted by nodes $s_1$ and $s_2$ conditioned on the value of the arithmetic sum. Indeed, note that the arithmetic sum of two i.i.d. Bernoulli random variables has a biased probability mass function and one can lower transmission rates by appropriately compressing these values. For zero-error compression in the single source setting, it is well recognized that variable length codes are needed. Accordingly, we consider the class of variable length network codes that allow for computation of the arithmetic sum. %Our converse arguments and achievable scheme both work in the variable length setting.

\subsection{Main contribution}
\begin{itemize}
\item We consider a variable-length network code for arithmetic sum computation in the particular DAG shown in Figure \ref{fig:fig1}. In this variable length setting, we present an upper bound and a lower bound for the computation rate. The upper bound arises from studying the entropy of the descriptions communicated by $s_1$ and $s_2$ to $t$ conditioned on the value of the sum. We show that this conditional entropy can be lower bounded by an appropriately characterized ``clumpy" distribution. The lower bound uses variable length codes for compression.
%\item For the case when the input data observed by the sources in Figure \ref{fig:fig1} are assumed to be independent iid and uniformly distributed over $\{0,1\}$, we give upper and lower bounds for computing capacity of the arithmetic sum. Due to the combinatorial nature of the problem, a special probability distribution called the clumpy distribution assumes importance in computing the upper bound for computing capacity.
\end{itemize}
This paper is organized as follows. Section \ref{sec:setup} presents the problem formulation. Section \ref{sec:upper_bd} uses a lower bound on the conditional entropy of the descriptions transmitted that is derived in Section \ref{sec:lb_entropy} to give an upper bound on the computation rate. Section \ref{sec:lower_bd} discusses an achievable scheme and Section \ref{sec:concl} concludes the paper. 
\section{Problem formulation}\label{sec:setup}
\begin{figure}[!t]
\centering
\includegraphics[width=2in]{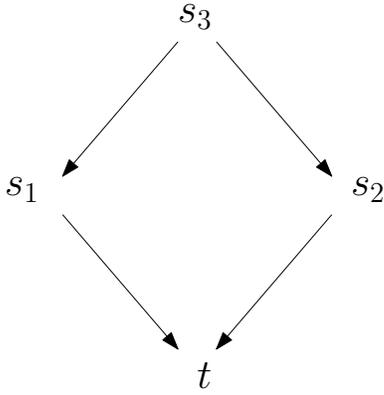}
% where an .eps filename suffix will be assumed under latex,
% and a .pdf suffix will be assumed for pdflatex; or what has been declared
% via \DeclareGraphicsExtensions.
\caption{A directed acyclic network with three sources and one terminal. \label{fig:fig1}}
\end{figure}
%We consider the problem of computing a function over the directed acyclic network in Fig \ref{fig:fig1}, where the directed edges (later denoted as an ordered pair of vertices) act as noiseless delay-free communication links.
The edges in Figure \ref{fig:fig1} (later denoted by an ordered pair of vertices) have unit-capacity. Suppose that $\mathcal{Z}$ is the alphabet used for communication, and $\mathcal{Z}>1$. $s_1, s_2, s_3$ are the three source nodes that observe independent uniform iid sources $X_1,X_2,X_3$ respectively, each from $\{0,1\}$. Terminal node $ t$ wants to compute the arithmetic sum $\Sigma=X_1+X_2+X_3, \Sigma \in \{0,1,2,3\}$. WLOG we assume that edges $(s_3,s_1), (s_3,s_2)$ forward the value of $X_3$ to $s_1,s_2$ respectively. We adapt a variable-length network code to this function computation problem. In what follows, all logarithms denoted as $\log$ are to the base $2$ unless specified otherwise. 
\begin{definition}
Let $\mathcal{Z}^\ast$ denote the set of all finite-length sequences with alphabet $\mathcal{Z}$. A variable-length $(k,N)$ network code for the network in Figure \ref{fig:fig1} has the following components.
\begin{enumerate}
\item Encoding functions for the edges, $e \in \{(s_1,t),(s_2,t)\}$:
\begin{IEEEeqnarray*}{c}
\phi_e:\{0,1\}^k\times \{0,1\}^k \rightarrow \mathcal{Z}^\ast
\end{IEEEeqnarray*}
Let $\bm{Z}_1:=\phi_{(s_1,t)}(\bm{X}_1,\bm{X}_3), \bm{Z}_2:=\phi_{(s_2,t)}(\bm{X}_2,\bm{X}_3)$ where $\bm{X}_j$ denotes a $k$-length random vector. $\bm{X}_j(i)$ refers to the $i$-th component of $\bm{X}_j$. We also let $\bm{X}_j^n$ represent the vector $(\bm{X}_j(1), \dots, \bm{X}_j(n))$.
\item Decoding function for the terminal $t$:
\begin{IEEEeqnarray*}{c}
\psi_t:\mathcal{Z}^N\times\mathcal{Z}^N \rightarrow \{0,1,2,3\}^k%
\end{IEEEeqnarray*}
where random variable $N$ is a \textit{stopping time} with respect to the sequence $(\bm{Z}_1(1),\bm{Z}_2(1)), (\bm{Z}_1(2),\bm{Z}_2(2)), \dots$. Thus, the indicator function $\bm{1}_{\{N=n\}}$ is a function of $((\bm{Z}_1(1),\bm{Z}_2(1),\hdots,(\bm{Z}_1(n),\bm{Z}_2(n))$.
\end{enumerate}
\end{definition}
%\begin{equation*}
%\bm{1}[N=n]=f((\bm{Z}_1(1),\bm{Z}_2(1)),\hdots,(\bm{Z}_1(n),\bm{Z}_2(n))),
%\end{equation*} where $\bm{1}[A]$ is the indicator function associated with event $\{A\}$ and $f:\mathcal{Z}^\ast\times\mathcal{Z}^\ast\rightarrow \bm{N}$ is a function specified by the network code.
%\begin{IEEEeqnarray*}{rL}
%\zeta &:\{0,1\}^\ast\times\{0,1\}^\ast \rightarrow \{0,1\},\\
%N &=\arg\min_{\tau \in \bm{N}}\zeta\left(\begin{IEEEeqnarraybox*}[][c]{,c/c/c/c,}
%\bm{Z}_1(1),&\bm{Z}_1(2),&\hdots ,&\bm{Z}_1(\tau),\\
%\bm{Z}_2(1),&\bm{Z}_2(2),&\hdots ,&\bm{Z}_2(\tau)
%\end{IEEEeqnarraybox*}\right)=1.%
%\end{IEEEeqnarray*}
%We let a random variable exponent denote an appropriately stopped sequence, i.e.,
%\begin{IEEEeqnarray*}{Rl}
%\bm{Z}_j^N:=&(\bm{Z}_j(1),\bm{Z}_j(2),\hdots,\bm{Z}_j(N)), \; j\in\{1,2\}.
%\end{IEEEeqnarray*}
Terminal $t$ estimates the $k$-length component-wise arithmetic sum (denoted as $\bm{\hat{\Sigma}}$) by setting $\bm{\hat{\Sigma}}:=\psi_t(\bm{Z}_1^N,\bm{Z}_2^N)$.%using the decoding and stopping functions as follows.
%\begin{IEEEeqnarray*}{C}
%\bm{\hat{\Sigma}}:=\psi_t(\bm{Z}_1^N,\bm{Z}_2^N).%
%\end{IEEEeqnarray*}
\begin{definition}
We say that a $(k,N)$ network code \textit{recovers} $\bm{\Sigma}$ \textit{with zero error} if $\Pr(\bm{\hat{\Sigma}}\neq\bm{\Sigma})=0,\;\text{for all}\;\bm{\Sigma} \in \{0,1,2,3\}^k$. The rate of such a network code is defined as $\frac{k}{\E N \log |\mathcal Z|}$, where $\E N$ denotes the expected value of $N$. The capacity is
\begin{equation*}
\mathcal{C}:=\sup \left\lbrace\frac{k}{\E N\log |\mathcal{Z}|}:~\begin{IEEEeqnarraybox*}[][c]{,t,} there is a zero-error $(k,N)$ \\network code that recovers $\Sigma$.
\end{IEEEeqnarraybox*} \right\rbrace
\end{equation*}
\end{definition}
%\begin{align*}
%\mathcal{C}&:=\sup \bigg{\{}\frac{k}{\E N \log |\mathcal Z|}: ~\text{there is a zero-error}\\
%& ~(k,N)~ \text{network code that recovers $\Sigma$.} \bigg{\}}
%\end{align*}
It can be observed that each component $1\leq i\leq k$ of $\bm{\Sigma}$ is independent and identically distributed as follows.
\begin{numcases}{\Pr(\bm{\Sigma}(i)=\beta) =}
1/8, & if $\beta\in \{0,3\}$,\nonumber\\
3/8, & if $\beta\in \{1,2\}$.\nonumber
\end{numcases}

%Note that our assumption on the distribution of the inputs $X_i, i \in \{1,2,3\}$ induces a probability distribution on $\bm{\Sigma}$ as follows.
%\begin{IEEEeqnarray*}{C}
%\Pr(\bm{\Sigma}=\bm{\sigma})=\prod_{i=1}^{k}\Pr(\bm{\Sigma}(i)=\bm{\sigma}(i)),
%\end{IEEEeqnarray*}
%where the probability mass function for each component is
%\begin{numcases}{\Pr(\bm{\sigma}(i))=}
%1/8, & if $\bm{\sigma}(i)\in \{0,3\}$,\nonumber\\
%3/8, & if $\bm{\sigma}(i)\in \{1,2\}$.\nonumber
%\end{numcases}
%If a particular $(k,N)$ network code recovers $\bm{\Sigma}$ with zero error, then the \textit{computation rate} for such a network code is equal to the ratio $k/\E N$ symbols per network use, where $\E N$ is the expected value of $N$.

%The above notion of zero error computation rate is related to the \textit{average case complexity} as defined by authors in [REF: Kowshik]. We can define the computing capacity of our network in a similar manner as done by authors in [REF:Appuswamy].
%\begin{definition}
%The arithmetic-sum computing capacity in bits per network use of the network in figure \ref{fig:fig1} is given below.
%\begin{equation*}
%\mathcal{C}:=\sup \left\lbrace\frac{k}{\E N}: ~\text{there is a zero-error} ~(k,N)~ \text{network code.}\right\rbrace
%\end{equation*}
%\end{definition}
In this work we derive upper and lower bounds on $\mathcal{C}$ for the network in Fig. \ref{fig:fig1}.
%\mathcal{C}:=\sup \left\lbrace\frac{k}{\E N}:~\begin{IEEEeqnarraybox*}[][c]{,t,} there is a zero-error \\$(k,N)$ network code.
%\end{IEEEeqnarraybox*} \right\rbrace
\section{Upper bound on computing capacity}
\label{sec:upper_bd}
Based on the relationships between the various quantities defined, we have the following.
\begin{IEEEeqnarray*}{rl}
&{}H(\bm{Z}_1^N,\bm{Z}_2^N,N)=H(\bm{Z}_1^N,\bm{Z}_2^N|N)+H(N),\\
{}&{}= H(N)+\sum\Pr(N=n)H(\bm{Z}_1^N,\bm{Z}_2^N|N=n),\\
{}&{}\leq H(N)+2\log |\mathcal{Z}|\E N.
\end{IEEEeqnarray*} The last inequality above is due to the fact that $H(\bm{Z}_1^N,\bm{Z}_2^N|N=n)\leq 2n\log |\mathcal{Z}|$ bits. Furthermore,
\begin{IEEEeqnarray*}{L}
H(N)+2\log|\mathcal{Z}|\E N \\
\geq H(\bm{Z}_1^N,\bm{Z}_2^N,N|\bm{\Sigma})+I(\bm{Z}_1^N,\bm{Z}_2^N,N;\bm{\Sigma}), \\
= H(\bm{\Sigma})-H(\bm{\Sigma}|\bm{Z}_1^N,\bm{Z}_2^N,N)+H(\bm{Z}_1^N,\bm{Z}_2^N,N|\bm{\Sigma}),\\
= H(\bm{\Sigma})+H(\bm{Z}_1^N,\bm{Z}_2^N,N|\bm{\Sigma}).
\end{IEEEeqnarray*} The last equality above is true by the zero-error criterion.
%because for all positive values of the probability mass $\Pr(\bm{Z}_1^N=\bm{Z}_1^N,\bm{Z}_2^N=\bm{Z}_2^N,N=n)$, we have that
%\begin{equation*}
%\Pr(\bm{\hat{\Sigma}}=\bm{\sigma}|\bm{Z}_1^N,\bm{Z}_2^N,n)=\Pr(\bm{\Sigma}=\bm{\sigma}|\bm{Z}_1^N,\bm{Z}_2^N,n)
%\end{equation*} by the zero error criterion. Noting that $\bm{\hat{\Sigma}}$ is a function of $(\bm{Z}_1^N,\bm{Z}_2^N,N)$ gives the result.
Note that the probability mass function for $\bm{\Sigma}$ implies that $H(\bm{\Sigma})=1.8113k$ bits. Thus, we have that
\begin{IEEEeqnarray}{c}
H(\bm{Z}_1^N,\bm{Z}_2^N,N|\bm{\Sigma})\leq H(N)+2\log|\mathcal{Z}|\E N-1.8113k.\IEEEeqnarraynumspace\label{eq:put_lb_entropy}%
\end{IEEEeqnarray}
Section \ref{sec:lb_entropy} derives a lower bound on the conditional entropy in the above inequality. Specifically, it is shown  that
\begin{IEEEeqnarray}{c}
H(\bm{Z}_1^N,\bm{Z}_2^N|\bm{\Sigma})\geq 0.75(-1+\log 3)k \;\text{bits for large}\; k. \IEEEeqnarraynumspace\label{eq:lb_entropy}
\end{IEEEeqnarray}
Using this in inequality \eqref{eq:put_lb_entropy}, we obtain
\begin{IEEEeqnarray}{Rl}
0.75(-1+\log 3)k \leq &H(N)+2\log|\mathcal{Z}|\E N-1.8113k, \IEEEnonumber \\
\implies \frac{k}{\E N \log |\mathcal{Z}|}\leq &\frac{H(N)}{\E N \log |\mathcal{Z}|}+\frac{2}{2.25}. \label{eq:put_lb_ratio}%
\end{IEEEeqnarray}
Furthermore, the following claim holds (see Appendix \ref{app:stopping_time_ratio} for a proof).
\begin{lemma}
For $k$ large enough, $H(N)/\E N \leq \epsilon$, for any $\epsilon> 0$.
\end{lemma}\label{lemma:hn_by en}
Indeed, for an arbitrary probability mass function on $\mathbb{N}$ (set of natural numbers), this ratio can take the value $1$ when $N$ follows a geometric distribution with parameter $1/2$. However, the zero error criterion restricts the possible set of probability mass functions for the stopping time $N$ and yields the required upper bound for the ratio.
\begin{lemma} \label{lemma:stopping_time}
A valid stopping time $N$ for our network satisfies
\begin{equation*}
\Pr(N=n)\leq \left(\frac{3}{8}\right)^k|\mathcal{Z}|^{2n} \;\text{for any}\; n\in \mathbb{N}.
\end{equation*}
\end{lemma}
\begin{IEEEproof} For a given $\bm{\sigma} \in \{0,1,2,3\}^k$ consider the set $S$ of values such that $\Pr(N = n|\bm{\Sigma}=\bm{\sigma})>0$. By definition of stopping time, terminal $t$ can recover at most $|\mathcal{Z}|^{2n}$ different values of $\bm{\hat{\Sigma}}$, which by the zero error criterion, is the same as the value of $\bm{\Sigma}$. Thus, if $|S|>|\mathcal{Z}|^{2n}$, there is a positive probability of error. Hence, we have,
\begin{IEEEeqnarray*}{Rl+x*}
\Pr(N=n)\leq &\sum_{\bm{\sigma}\in S} \Pr(N= n|\bm{\Sigma}=\bm{\sigma})\Pr(\bm{\Sigma}=\bm{\sigma}),\\
\leq &|\mathcal{Z}|^{2n}\max_{\sigma}\Pr(\bm{\Sigma}=\bm{\sigma}),\\
= &\left(\frac{3}{8}\right)^k|\mathcal{Z}|^{2n}. %&\IEEEnonumber\IEEEQEDhere
\end{IEEEeqnarray*}
\end{IEEEproof}

\section{Lower bound on conditional entropy}\label{sec:lb_entropy}
In this section we derive the lower bound on $H(\bm{Z}_1^N,\bm{Z}_2^N|\bm{\Sigma})$ as stated in inequality \eqref{eq:lb_entropy}. To do this, we first note that the zero error criterion enforces a requirement that the stopped sequences $\bm{Z}_1^N, \bm{Z}_2^N$ must satisfy.
\begin{lemma}\label{lemma:labels}
For a valid $(k,N)$ network code, let $\bm{z}_1^{n_1}:=\phi_{(s_1,t)}(\bm{x}_1, \bm{x}_3)$ and $\bm{z}_1^{n_1'}:=\phi_{(s_1,t)}(\bm{x}_1',\bm{x}_3)$. Similarly define $\bm{z}_2^{n_2}$ and $\bm{z}_2^{n_2'}$. Then,% for any $\bm{x}_3 \in \{0,1\}^k$,
\begin{itemize}
\item $\bm{z}_1^{n_1} \neq \bm{z}_1^{n_1'}$ for all $\bm{x}_1 \neq \bm{x}_1'$ with $\bm{x}_1, \bm{x}_1' \in \{0,1\}^k$ and
\item $\bm{z}_2^{n_2} \neq \bm{z}_2^{n_2'}$ for all $\bm{x}_2 \neq \bm{x}_2'$ with $\bm{x}_2, \bm{x}_2' \in \{0,1\}^k$.
\end{itemize}
\end{lemma}
\begin{IEEEproof}
Assume otherwise and consider the two sets of inputs $(\bm{x}_1,\bm{x}_2,\bm{x}_3)$ and $(\bm{x}_1',\bm{x}_2,\bm{x}_3)$ such that $\bm{x}_1'+\bm{x}_2+\bm{x}_3 \neq \bm{x}_1+\bm{x}_2+\bm{x}_3$. One can easily see that such a set of inputs exist. Then if $\bm{z}_1^{n_1}=\bm{z}_1^{n_1'}$, the terminal $t$ is unable to compute the arithmetic sum correctly from the corresponding stopped sequences, leading to a non zero probability of error.
\end{IEEEproof}

%\textbf{COMMENT:} I think the above lemma doesn't generalize for $\epsilon$-error case. To see this, a trivial scheme for $Z_1$ label that doesn't satisfy the condition in the lemma is described and it is shown that the error goes to $0$ as $k$ increases. Suppose $a,b \in \{0,1\}^k$ are two realisations of $X_1$ such that $Z_1(a,x_3)=Z_1(b,x_3)$. Suppose $Z_1$ assigns a different label to all other $(X_1,X_3)$ pairs and suppose $Z_2$ assigns a different label to each $(X_2,X_3)$ pair. Then the probability of error is not more than $2/8^k$ which goes to $0$ for large $k$.
Set $L_{x,y}:=2^{x+y}$ and $M_{x,y}:=3^{x+y}$. For a natural number $u$, let $[u]:=\{1,2,\hdots,u\}$. For a vector $\bm{v}$, index its components with a natural number $i$ and let $\bm{v}(i)$ denote its $i$-th component.
\begin{claim}
Let a particular realization $\bm{\sigma}$ of $\bm{\Sigma}$ be such that $x$ components of it equal $1$ and $y$ components of it equal $2$. For a valid $(k,N)$ network code, the conditional entropy $H(\bm{Z}_1^N,\bm{Z}_2^N|\bm{\Sigma}=\bm{\sigma})$ is minimized when the probability mass function $\Pr(\bm{Z}_1^N=\bm{z}_1^N,\bm{Z}_2^N=\bm{z}_2^N|\bm{\Sigma}=\bm{\sigma})$ is positive for exactly $L_{x,y}$ distinct $(\bm{z}_1^N,\bm{z}_2^N)$ pairs.
\end{claim}
\begin{IEEEproof}
For a $\bm{\sigma}$ with $x$ 1's and $y$ 2's in it, there are $M_{x,y} ~(\bm{x}_1,\bm{x}_2,\bm{x}_3)$-tuples that result in that particular sum. Within these $M_{x,y}$ input tuples, there are $L_{x,y}$ different values of $\bm{x}_3$. We can partition all input tuples into disjoint sets that have the arithmetic sum $\bm{\sigma}$ based on the value of $\bm{x}_3$. The set with the most number of input tuples in it corresponds to a particular value which we denote $\bm{\tilde{x}}_3$. One can check that, for $i=\{1,2,\hdots,k\}$
\begin{equation*}
\setlength{\nulldelimiterspace}{0pt}
\bm{\tilde{x}}_3(i)=\left\{\begin{IEEEeqnarraybox}[\relax][c]{l's}
0,&if $\sigma(i)=0,1$\\
1,&if $\sigma(i)=2,3.$%
\end{IEEEeqnarraybox}\right.
\end{equation*}
From Lemma \ref{lemma:labels}, we know that all input tuples for a fixed $\bm{x}_3$  must receive distinct $(\bm{Z}_1^N,\bm{Z}_2^N)$ labels. Thus for any $\bm{\sigma}$, we must have atleast $L_{x,y}$ distinct $(\bm{z}_1^N,\bm{z}_2^N)$ labels as the size of the largest partition (which is the $\bm{\tilde{x}}_3$-partition) is $L_{x,y}$. These instantiations of the pair process $(\bm{Z}_1^N,\bm{Z}_2^N)$ must therefore have a positive conditional probability. Note that all the input tuples that result in a particular $\bm{\sigma}$ are equally likely. Hence, the conditional probability of any one particular $(\bm{z}_1^N,\bm{z}_2^N)$ label for a given $\bm{\sigma}$ has to be a multiple of $1/M_{x,y}$.

In Appendix \ref{app:support_arg} we prove the following claim from which the result follows. Let $c > 0$ and $u^*$ be a positive integer such that $c u^* \leq 1$. For a natural number $u \leq u^*$, let $\mathcal{Q}_{u}$ be the set of probability mass functions supported on $[u]$ such that for all vectors $\bm{q} \in \mathcal{Q}_u$, we have that $\bm{q}(i)\geq c>0, ~\forall i \in [u]$.
\begin{claim}\label{claim:entropy_support_set}
For some $m \leq u^*-1$, let
\begin{IEEEeqnarray*}{Rl}
\bm{q}_{m}:=\arg\min_{\bm{q} \in \mathcal{Q}_m}H(\bm{q}), \;\text{and}\; \bm{q}_{m+1}:=\arg\min_{\bm{q} \in\mathcal{Q}_{m+1}}H(\bm{q}).
\end{IEEEeqnarray*}
Then, we have $H(\bm{q}_m) \leq H(\bm{q}_{m+1})$.
\end{claim}
This follows from the fact that entropy is a concave function and attains its minimum at an extremal point of the underlying polyhedron. Thus, using any more than $L_{x,y}$ distinct labels will increase the conditional entropy.
\end{IEEEproof}

We now explicitly derive the conditional entropy-minimizing distribution over $L_{x,y} ~(\bm{z}_1^N,\bm{z}_2^N)$ labels. Index all the $L_{x,y}$ different values of $\bm{x}_3$ that result in a particular arithmetic sum $\bm{\sigma}$ by a natural number $i$ and denote them as $\bm{x}_3^i, i \in \{1,2,\hdots,L_{x,y}\}$. Recall that the input tuples with sum $\bm{\sigma}$ can be partitioned into disjoint sets based on the value of $\bm{x}_3^i$. We call the set corresponding to $\bm{x}_3^i$, the $\bm{x}_3^i$-partition. Let $n_i$ be the size of $\bm{x}_3^i$-partition.
\begin{claim}\label{claim:family}
Conditioned on the particular value of $\bm{\sigma}$, the set of all probability mass functions on $L_{x,y}$ valid $(\bm{z}_1^N,\bm{z}_2^N)$ labels can be represented by a family $\mathcal{P}$ of vectors over the reals.
\begin{IEEEeqnarray}{l}
\setlength{\nulldelimiterspace}{0pt}
\mathcal{P}=\left\{\begin{IEEEeqnarraybox}[\relax][c]{l}
\bm{p} \in \mathbb{R}^{L_{x,y} \times 1}: \bm{p}=\frac{1}{M_{x,y}}\sum_{i=1}^{L_{x,y}}\bm{e}_i, \bm{e}_i \in \mathbb{R}^{L_{x,y} \times 1}\\
\text{such that}~ \bm{e}_i ~\text{has}~ n_i ~\text{1's and rest as 0 for every}~ i.%
\end{IEEEeqnarraybox}\right\}\label{eq:family}\IEEEeqnarraynumspace%
\end{IEEEeqnarray}
\end{claim}
\begin{IEEEproof}
Each $\bm{x}_3^i$-partition of the $(\bm{x}_1,\bm{x}_2,\bm{x}_3)$ input tuples that have the arithmetic sum $\bm{\sigma}$ has $n_i$ elements in it.
%For any value of the arithmetic sum $\bm{\sigma}$, there are $L_{x,y}$ distinct $\bm{X_3^i}$-partitions. Each of these $\bm{X_3^i}$-partitions in general has a different number of input tuples in it.
Each equally-likely input tuple for this $\bm{\sigma}$ is assigned a particular $(\bm{z}_1^N,\bm{z}_2^N)$ label by the encoding functions. Thus, the conditional probability distribution $\Pr(\bm{Z}_1^N=\bm{z}_1^N,\bm{Z}_2^N=\bm{z}_2^N|\Sigma=\sigma)$ is determined by how frequently the $(\bm{z}_1^N,\bm{z}_2^N)$ label is reassigned to different input tuples that have the arithmetic sum as $\sigma$.
%Index all the different values of $\bm{X}_3$ (there are a total of $L_{x,y}$) that result in a particular arithmetic sum $\bm{sigma}$ by a natural number $i$ and denote them as $\bm{X_3^i}, i \in \{1,2,\hdots,L_{x,y}\}$. Let $n_i:=|\bm{X_3^i}|$.

Define a $L_{x,y}$-length vector $\bm{e}_i$ for each $\bm{x}_3^i$-partition. The components of the vector $\bm{e}_i$ denote whether a particular $(\bm{z}_1^N,\bm{z}_2^N)$ label is assigned to an input tuple in the $\bm{x}_3^i$-partition or not. Note that by Lemma \ref{lemma:labels}, a $(\bm{z}_1^N,\bm{z}_2^N)$ label can be assigned to atmost one input tuple in a particular $\bm{x}_3^i$-partition. Thus $\bm{e}_i$ has $n_i$ components as $1$ and the rest as $0$. Then the frequency of occurrence of a particular $(\bm{z}_1^N,\bm{z}_2^N)$ label among all the input tuples that result in the arithmetic sum $\bm{\sigma}$ can be found by considering the component-wise sum $\sum_{i=1}^{L_{x,y}}\bm{e}_i$. Normalizing by the total number of input tuples gives the claim.
%\begin{equation*}
%\sum_{i=1}^{L_{x,y}}\bm{e}_i.
%\end{equation*}
\end{IEEEproof}
%There are $L_{x,y}$ distinct $\bm{X}_3$-partitions for any value of $\bm{\sigma}$, and let $n_i, i=\{1,2,\hdots,L_{x,y}\}$ denote the number of input tuples in the $i$-th $\bm{X}_3$-partition.
%\begin{align} \label{eq:family}
%\mathcal{P}=\left\lbrace p \in \bm{R}^{L_{x,y} \times 1} : p=\frac{1}{M_{x,y}}\sum_{i=1}^{L_{x,y}}e_i ~\text{where}~ e_i ~\text{has}~ n_i ~1's~\text{and rest}~0's.\right\rbrace
%\end{align}
%Indexing all the different values of $\bm{X}_3$ that result in a particular $\bm{\sigma}$ as $\bm{X_3^i}, i=\{1,2,\hdots,L_{x,y}\}$ we see that each $e_i$ vector of length $L_{x,y}$ represents the different $(\bm{Z}_1^N,\bm{Z}_2^N)$ labels assigned to the $(\bm{X}_1,\bm{X}_2,\bm{X_3^i})$ tuples present in the $i$-th $\bm{X}_3$-partition of the input space for that particular $\bm{\sigma}$.
\begin{theorem}
Let $\bm{1}_u$ and $\bm{0}_v$ denote the all-ones vector with $u$ components and the all-zeros vector with $v$ components respectively. Let $L_{x,y}^i:=L_{x,y}-n_i$ for all $i \in [L_{x,y}]$. Then
\begin{equation}
\bm{p}^\star=\frac{1}{M_{x,y}}\left(\begin{bmatrix}
\mathbf{1}_{n_1}\\
\mathbf{0}_{L_{x,y}^1}
\end{bmatrix}+
\begin{bmatrix}
\mathbf{1}_{n_2}\\
\mathbf{0}_{L_{x,y}^2}
\end{bmatrix}+\hdots +\begin{bmatrix}
\mathbf{1}_{n_{L_{x,y}}}\\
\mathbf{0}_{L_{x,y}^{L_{x,y}}}
\end{bmatrix}\right) \label{eq:pstar}
\end{equation}
%\begin{IEEEeqnarray}{rCl}
%\IEEEeqnarraymulticol{3}{c}{\bm{p}^\star =\frac{A}{M_{x,y}},}\IEEEyesnumber\IEEEyessubnumber\\
%\noalign{\noindent with\vspace{\jot}}
%A&=&\begin{bmatrix}
%\mathbf{1}_{n_1}\\
%\mathbf{0}_{L_{x,y}-n_1}
%\end{bmatrix}+
%\begin{bmatrix}
%\mathbf{1}_{n_2}\\
%\mathbf{0}_{L_{x,y}-n_2}
%\end{bmatrix}+\hdots +\begin{bmatrix}
%\mathbf{1}_{n_{L_{x,y}}}\\
%\mathbf{0}_{L_{x,y}-n_{L_{x,y}}}
%\end{bmatrix}\IEEEyessubnumber\IEEEeqnarraynumspace\label{eq:pstar}%
%\end{IEEEeqnarray}
is a probability mass function in $\mathcal{P}$ that minimizes the entropy on $L_{x,y}$ $(\bm{z}_1^N,\bm{z}_2^N)$ labels. Moreover, any other entropy-minimizing distribution is only a permutation of $\bm{p}^\star$.
%, i.e., the set of components of a vector denoting an entropy-minimizing distribution is equal to the set of components of $\bm{p}^\star$.
\end{theorem}
\begin{IEEEproof}
It will be shown in Claim \ref{claim:extremal_point} that $\bm{p}^\star$ is an an extremal point of the set conv$(\mathcal{P})$, which is the convex hull of the set $\mathcal{P}$. Thus, it is a potential minimizer of the concave entropy function over the convex set conv$(\mathcal{P})$. Claim \ref{claim:convex_comb} shows that there are no other candidate minimizers, and hence
\begin{equation*}
H(\bm{Z}_1^N,\bm{Z}_2^N|\bm{\Sigma}=\bm{\sigma})\geq \min_{\bm{p} \in \mathcal{P}}H(\bm{p}) \geq \min_{\bm{p} \in \text{conv}(\mathcal{P})}H(\bm{p})=H(\bm{p}^\star).
\end{equation*}
We refer to $\bm{p}^\star$ as the ``clumpy" distribution.
\end{IEEEproof}
%\begin{align} \label{eq:pstar}
%p^\star=\frac{1}{M_{x,y}}\left( \begin{bmatrix}
%\mathbf{1}_{n_1}\\
%\mathbf{0}_{L_{x,y}-n_1}
%\end{bmatrix}+
%\begin{bmatrix}
%\mathbf{1}_{n_2}\\
%\mathbf{0}_{L_{x,y}-n_2}
%\end{bmatrix}+ \hdots+
%\begin{bmatrix}
%\mathbf{1}_{n_{L_{x,y}}}\\
%\mathbf{0}_{L_{x,y}-n_{L_{x,y}}}
%\end{bmatrix}\right)
%\end{align}

Let $\bfe_i^\star=\begin{bmatrix}
\mathbf{1}_{n_i} ~ \mathbf{0}_{L_{x,y}^i}
\end{bmatrix}^\intercal$ so that $\bfp^\star = \frac{1}{M_{x,y}}\sum_{i=1}^{L_{x,y}}\bfe_i^\star$. Let $\bfe_i$ denote any binary vector of length $L_{x,y}$ such that it has exactly $n_i$ ones. From Claim \ref{claim:family} any $\bfp \in \mathcal{P}$ can be expressed as $\sum_{i=1}^{L_{x,y}} \bfe_i$ for appropriate choices of vectors $\bfe_i, i \in [L_{x,y}]$.
\begin{claim}\label{claim:t}
Let $\bfd = \bfp^\star - \bfp$ and let $\bfd(i)$ represent its $i$-th component. Then, $\sum_{i=1}^u \bfd(i) \geq 0$ for all $u \in [L_{x,y}]$.
\end{claim}
\begin{IEEEproof}
We show this by considering $\bfe_j^\star - \bfe_j$. Note that, for $1 \leq i \leq n_i$, we have $\bfe_j^\star(i) = 1 \geq \bfe_j(i)$. This implies that for $1 \leq u \leq n_i$, we have
\begin{align*}
\sum_{i=1}^u \bfe_j^\star(i) - \bfe_j(i) &\geq 0.
\end{align*}
On the other hand when $n_i < u \leq L_{x,y}$, we have
\begin{align*}
\sum_{i=1}^u \bfe_j^\star(i) - \bfe_j(i) & = n_i - \sum_{i=1}^z \bfe_j(i) \geq 0,
\end{align*}
where the last inequality holds because both $\bfe_j^\star$ and $\bfe_j$ have $n_i$ ones. As $\bfd = \frac{1}{M_{x,y}}\sum_{j=1}^{L_{x,y}} (\bfe_j^\star - \bfe_j)$, the result follows.
\end{IEEEproof}

\begin{corollary}
If $\bm{p} \in \text{conv}(\mathcal{P})$ and $\bm{p} \neq \bm{p}^\star$, then
\begin{itemize}
\item $\bm{p}^\star(i) > \bm{p}(i)$  for $i=\text{min}\{k: \bm{p}^\star(k)\neq \bm{p}(k)\}$, and
\item $\bm{p}^\star(j) < \bm{p}(j)$  for $j=\text{max}\{k: \bm{p}^\star(k)\neq \bm{p}(k)\}$.
\end{itemize}
\end{corollary}
\begin{IEEEproof}
For $\bm{p} \in \mathcal{P}$, substitute $u=\text{min}\{i: \bm{p}^\star(i)\neq \bm{p}(i)\}$ in Claim \ref{claim:t} and note that $\bm{d}(i)=0 ~\forall i<u$. For the second case, note that $\sum_{i=1}^{\text{max}\{j: \bm{p}^\star(j)\neq \bm{p}(j)\}}\bm{d}(i)=0$ and substitute $u=\text{max}\{j: p^\star(j)\neq p(j)\}-1$ in claim \ref{claim:t}.

Now let $\bm{p} \in \text{conv}(\mathcal{P})$ such that $\bm{p}=\sum \mu_l \bm{p}_l$ where each $\bm{p}_l \in \mathcal{P}$ and each $\mu_l > 0$ with $\sum \mu_l=1$. Since the corollary is true for each $\bm{p}_l$, we have that $\bm{p}_l(i_l)<\bm{p}^\star(i_l)$, if $i_l$ is the first index where $\bm{p}_l$ differs from $\bm{p}^\star$. Suppose $i_j:=\min_l i_l$ is unique, then $\bm{p}_l(i_j)=\bm{p}^\star(i_j) ~\forall l \neq j$ and $\bm{p}_j(i_j)<\bm{p}^\star(i_j)$. Hence $\bm{p}(i_j)=(1-\mu_j)\bm{p}^\star(i_j)+\mu_j \bm{p}_j(i_j) < \bm{p}^\star(i_j)$. A similar argument also works when $i_j$ is not unique.
\end{IEEEproof}
%Then we can make the following claim.
\begin{claim}\label{claim:extremal_point}
$\bm{p}^\star$ is an extremal point of the convex set conv($\mathcal{P}$).
\end{claim}
\begin{IEEEproof}
Suppose that there exist $\bm{p}_1,\bm{p}_2 \in \text{conv}(\mathcal{P})$ such that $p^\star=(\bm{p}_1+\bm{p}_2)/2$ with $\bm{p}_1 \neq \bm{p}_2$.
Let $i$ be the least index such that $\bm{p}_1(i) \neq \bm{p}_2(i)$. Then without loss of generality, $\bm{p}_1(i)>\bm{p}^\star(i)>\bm{p}_2(i)$. But that is a contradiction to the corollary to Claim \ref{claim:t} and hence our claim is proved.
\end{IEEEproof}
Our next claim shows that {\it all} extremal points of conv($\mathcal{P}$) are permutations of the distribution $\bm{p}^\star$ (proof appears in Appendix \ref{app:convex_comb}).
\begin{claim}\label{claim:convex_comb}
Any $\bm{p} \in$ conv($\mathcal{P}$) can be written as a convex combination of vectors which are permutations of $\bm{p}^\star$.
\end{claim}
%\begin{IEEEproof}
%See Appendix \ref{app:convex_comb}.
%\end{IEEEproof}
\subsection{Entropy of clumpy distribution}\label{subsec:clumpy_entropy}
\begin{figure}
\centering
\includegraphics[scale=0.8]{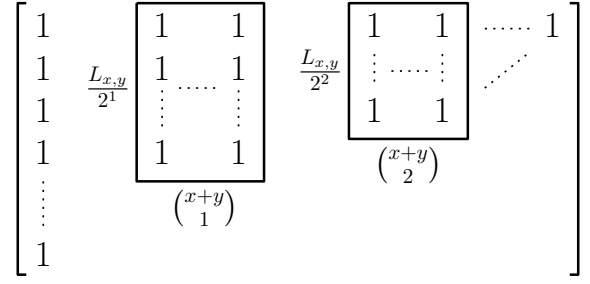}
\caption{Staircase structure for the $\{0,1\}$-matrix $\bm{A}$ defined in subsection \ref{subsec:clumpy_entropy}. Only the $1$s are shown. Boxes denote all-ones block matrices of row and column dimension as mentioned beside their length and breadth.}\label{fig:A}
\end{figure}

For $\bm{p}^\star$ as defined in \eqref{eq:pstar}, WLOG assume that $n_1 \geq n_2 \geq \hdots \geq n_{L_{x,y}}$. Recall that $n_i$ is the number of input tuples of the form $(\bm{x}_1,\bm{x}_2,\bm{x}_3^i)$ that have the arithmetic sum $\bm{\sigma}$. Then $n_1=L_{x,y}$ as it corresponds to $\bm{\tilde{x}}_3$-partition. If a particular $\bm{x}_3$ differs from $\bm{\tilde{x}}_3$ in any $1 \leq u \leq x+y$ components, then one can check that the number of input tuples with arithmetic sum $\bm{\sigma}$ in this particular $\bm{x}_3$-partition is exactly $L_{x,y}/2^u$. Also, depending on which $u$ bits of $\bm{\tilde{x}}_3$ are flipped, there are $\binom{x+y}{u}$ different $\bm{x}_3$-partitions that have $L_{x,y}/2^u$ input tuples in them. Let $\bm{A}$ denote a $L_{x,y}\times L_{x,y}$ matrix with the $i$th column as the vector $\bm{e}_i^\star$ for all $i \in [L_{x,y}]$. By the above discussion, matrix $\bm{A}$ has a ``staircase" structure as shown in Figure \ref{fig:A}.
Each row of $\bm{A}$ corresponds to a particular label and the row sum indicates its frequency of occurrence. Conditioned on $\bm{\sigma}$, each input tuple that results in that sum is equally likely with probability $1/M_{x,y}$. Then the expression for the entropy of the clumpy distribution is given in equation \eqref{eq:clumpy_entropy}. The first term in the RHS denotes the contribution to the entropy by labels which are repeated exactly once, the second term corresponds to labels repeated exactly twice, and so on.
\begin{IEEEeqnarray*}{Rl}
H(\bm{p}^\star)=&-2^{x+y-1}\frac{\binom{x+y}{0}}{M_{x,y}}\log\frac{\binom{x+y}{0}}{M_{x,y}}\\&-2^{x+y-2}\frac{\binom{x+y}{0}+\binom{x+y}{1}}{M_{x,y}}\log\frac{\binom{x+y}{0}+\binom{x+y}{1}}{M_{x,y}}-\hdots\\
&-\frac{\binom{x+y}{0}+\hdots+\binom{x+y}{x+y}}{M_{x,y}}\log\frac{\binom{x+y}{0}+\hdots+\binom{x+y}{x+y}}{M_{x,y}}\IEEEyesnumber \IEEEeqnarraynumspace \label{eq:clumpy_entropy}
\end{IEEEeqnarray*}

%We have that $H(\bm{Z}_1^N,\bm{Z}_2^N)=1.8113k+H(\bm{Z}_1^N,\bm{Z}_2^N|\bm{\Sigma})$.
Let the set of $(\bm{z}_1^N,\bm{z}_2^N)$ labels assigned to the input tuples for a particular value of $\bm{\Sigma}=\bm{\sigma}$ be $Z_{\bm{\sigma}}$. Then our discussion about the clumpy distribution implies that a lower bound for the quantity $H(\bm{Z}_1^N,\bm{Z}_2^N|\bm{\Sigma}=\bm{\sigma})$ can be obtained by choosing $|Z_{\bm{\sigma}}|=L_{x,y}$ and letting these labels follow the probability mass function $\bm{p}^\star$.
Since the network has to compute the arithmetic-sum of the messages, input tuples that result in a different $\bm{\sigma}$ must be provided a different $(\bm{z}_1^N,\bm{z}_2^N)$ label. Thus, for two different realizations $\bm{\sigma},\bm{\tilde{\sigma}}$ that have the same number of 1's and 2's, we have that $Z_{\bm{\sigma}} \cap Z_{\bm{\tilde{\sigma}}}=\phi$ and both $H(\bm{Z}_1^N,\bm{Z}_2^N|\bm{\Sigma}=\bm{\sigma}),H(\bm{Z}_1^N,\bm{Z}_2^N|\bm{\Sigma}=\bm{\tilde{\sigma}})\geq H(\bm{p}^\star)$, where $x,y$ in the definition (equation \eqref{eq:pstar}) of $\bm{p}^\star$ are the number of 1's, 2's respectively in $\bm{\sigma}$ or $\bm{\tilde{\sigma}}$.

%The number of different equally likely input tuples that result in a particular value of $\sigma$ are $M_{x,y}$.

%= &\frac{1}{M_{x,y}}\sum_{j=1}^{x+y+1}\lceil 2^{x+y-j}\rceil \sum_{s=0}^{j-1}\binom{x+y}{s}B(x,y,j)  \IEEEyesnumber \IEEEeqnarraynumspace \label{eq:lower_bd_expr_A}
%\end{IEEEeqnarray*}
%where
%\begin{equation*}
%B(x,y,j)=(x+y)\log 3-\log \sum_{t=0}^{j-1}\binom{x+y}{t}.
%\end{equation*}
% Note that $H(\bm{Z}_1^N,\bm{Z}_2^N|\bm{\Sigma})=\sum_{\bm{\sigma}}\Pr(\bm{\Sigma}=\bm{\sigma})H(\bm{Z}_1^N,\bm{Z}_2^N|\bm{\Sigma}=\bm{\sigma})$.
Thus a lower bound for $H(\bm{Z}_1^N,\bm{Z}_2^N|\bm{\Sigma})=\sum_{\bm{\sigma}}\Pr(\bm{\Sigma}=\bm{\sigma})H(\bm{Z}_1^N,\bm{Z}_2^N|\bm{\Sigma}=\bm{\sigma})$ can be found by assuming that each of the conditional pmfs are clumpy. Using this, in Appendices \ref{app:clumpy_entropy} and \ref{app:average_clumpy_entropy}, we show the following result.
\begin{lemma}\label{lemma:limit_H_by_k}
\begin{align*}
\frac{H(\bm{Z}_1^N,\bm{Z}_2^N|\Sigma)}{k} &{}\rightarrow 0.75(-1+\log 3) \;\text{as}\; k\rightarrow \infty.
\end{align*}
\end{lemma}

\section{Lower bound on computing capacity}\label{sec:lower_bd}
In this section we describe a valid $(k,N)$ network code which satisfies $k/\E N=2/2.5$ for the case when $\mathcal{Z}=\{0,1\}$. A similar scheme can be easily extended to larger alphabets.
In fact, the scheme described here is the same scheme as the one described in \cite{appuFKZ11}, except that no probability distribution on the inputs was used there. In what follows, all addition operations are over the real numbers.

Set $k$ to be an even number. The encoder at $s_1$ computes the first $k/2$ components of the sum $\bm{X}_1+\bm{X}_3$. Note that the components of $\bm{X}_1+\bm{X}_3$ are iid distributed according to
\begin{numcases}{\Pr(X_1+X_3=u)=}
1/4,& if $u \in \{0,2\}$ \nonumber\\
1/2,& otherwise, \nonumber
\end{numcases} and hence $H(X_1+X_3)=1.5$ bits.
For an $\epsilon >0$, we compress the sequence of first $k/2$ components of $\bm{X}_1+\bm{X}_3$ based on whether they belong to the weakly typical set $A_\epsilon^{(k/2)}$ \cite{coverthomas} or not. We can encode all the sequences in $A_\epsilon^{(k/2)}$ by using atmost $\lceil \frac{k}{2}(1.5+\epsilon)\rceil$ bits. For encoding sequences not in $A_\epsilon^{(k/2)}$, we don't need more than $\lceil \frac{k}{2}\log 3\rceil$ bits. We add an extra bit to indicate whether the sequence being encoded belongs to the typical set or not. Having encoded the first $k/2$ bits of $\bm{X}_1+\bm{X}_3$ in the above fashion, $s_1$ transmits the subsequent $k/2$ bits of $\bm{X}_1$ in an uncoded manner.

The encoder at $s_2$ employs a similar procedure as above except that it transmits the first $k/2$ bits of $\bm{X}_2$ in an uncoded manner and the subsequent $k/2$ bits of the component-wise sum $\bm{X}_2+\bm{X}_3$ using typical set coding for a typical set with the same $\epsilon$.

The terminal is able to recover the first $k/2$ components of $\bm{X}_1+\bm{X_3}$ and $\bm{X}_2$ with zero error and the last $k/2$ components of $\bm{X}_1$ and $\bm{X}_2+\bm{X}_3$ with zero error. From these it can correctly compute $k$ components of the sum $\bm{X}_1+\bm{X}_2+\bm{X}_3$.

For the value of the stopping time, the terminal waits for $1+k/2$ bits so as to obtain all the uncoded bits and the information about whether the coded bits belong to the typical set or not. Based on that, it waits for an appropriate number of bits so as to decode the required information without error. Let $\bm{V}_1,\bm{V}_2$ denote the first $k/2$ components of $\bm{X}_1+\bm{X}_3$ and the last $k/2$ components of $\bm{X}_2+\bm{X}_3$ respectively. Let $B$ denote the event $\{\bm{V_1}\in A_\epsilon^{(\frac{k}{2})} \cap \bm{V}_2 \in A_\epsilon^{(\frac{k}{2})}\}$. Then the expected value of the stopping time can be evaluated as follows
\begin{IEEEeqnarray*}{Rl}
\E N =&1+\frac{k}{2}+\Pr(B)\left\lceil \frac{k}{2}(1.5+\epsilon)\right\rceil +(1-\Pr(B))\left\lceil \frac{k}{2}\log 3 \right\rceil ,\\
\leq & 1+\frac{k}{2}+\left\lceil \frac{k}{2}(1.5+\epsilon)\right\rceil + 2\epsilon \left\lceil \frac{k}{2}\log 3\right\rceil.
\end{IEEEeqnarray*}
Hence, for large $k$, $\E N \approx \frac{5k}{4}$ and that gives our result.

\section{Conclusions and future work}
\label{sec:concl}
We have obtained new upper and lower bounds for zero error arithmetic-sum computation using variable-length network codes for a specific network. There is still a gap between the achievable rate and the upper bound. Future work will involve trying to narrow this gap. In addition, all currently known upper bounds for function computation over DAGs are based on cutsets and are recognized to be loose. It may be fruitful to examine whether the upper bound technique used in this work that operates by lower bounding the entropy of the descriptions conditional on the function value are applicable in more general scenarios.

%The encoding scheme described does not match the upper bound and future work would involve closing that gap. We would also look at bounds given by employing the clumpy distribution for other problem instances of function computation.


%\section*{Acknowledgment}



\bibliographystyle{IEEEtran}
\bibliography{IEEEabrv,../BibFiles/tip}
%\begin{thebibliography}{1}
%\end{thebibliography}
%\bibliographystyle{IEEEtran}
%\bibliography{../FunCompBound/tip}
\appendices
\section{}\label{app:stopping_time_ratio}
Consider the optimization problem defined as follows, where $p_i:=\Pr(N=i)$ for $i \in \mathbb{N}$ and $\Delta=\log e/\epsilon$.
\begin{IEEEeqnarray}{uL}
minimize\hspace{15pt} & \E N-\frac{\Delta}{\log e} H(N) = \sum p_i(i+\Delta\ln p_i) \IEEEyesnumber\IEEEyessubnumber\IEEEeqnarraynumspace \label{opt:min_obj}\\
subject to:\hspace{10pt} & p_i-\left(\frac{3}{8}\right)^k|\mathcal{Z}|^{2i} \leq 0 \;\text{for all}\; i \in \mathbb{N}, \IEEEyessubnumber \label{opt:lemma1}\\
& -p_i \leq 0 \;\text{for all}\; i \in \mathbb{N},\IEEEnonumber*\\
& \sum p_i -1 =0.
\end{IEEEeqnarray}
The Lagrangian of the objective function \eqref{opt:min_obj} is
\begin{IEEEeqnarray}{Rl}
L(\bm{p},\lambda,\bm{\nu},\bm{\mu})=&\sum p_i(i+\Delta\ln p_i)+\lambda (1-\sum p_i)\IEEEnonumber\\&-\sum \bm{\nu}_i p_i + \sum \bm{\mu}_i\left(p_i-\left(\frac{3}{8}\right)^k|\mathcal{Z}|^{2i}\right).\IEEEeqnarraynumspace \label{eq:lagrangian}%
\end{IEEEeqnarray}
Here, $\lambda, \bm{\nu}\geq \bm{0}, \bm{\mu}\geq \bm{0}$ are dual variables with the natural number subscript $i$ indexing their components. Since the Lagrangian is convex in $\bm{p}$, minimizing it involves setting
\begin{IEEEeqnarray*}{Rl}
\frac{\partial L}{\partial p_i}=&i+\Delta\ln p_i+\Delta - \lambda - \nu_i +\mu_i=0 ~\text{for all}~i,\\
\implies p_i=&\exp\left(\frac{\lambda + \nu_i - \mu_i -i -\Delta}{\Delta}\right).
\end{IEEEeqnarray*}
Substituting this back in equation \eqref{eq:lagrangian}, we get that the dual function is
\begin{IEEEeqnarray*}{C}
\mathcal{L}(\lambda,\bm{\mu},\bm{\nu})=\lambda-\sum_i B(i,\lambda,\mu_i,\nu_i),
\end{IEEEeqnarray*}
where
\begin{IEEEeqnarray*}{Rl}
B(i,\lambda,\mu_i,\nu_i):=&\left[\Delta \exp\left(\frac{\lambda + \nu_i - \mu_i -i -\Delta}{\Delta}\right)\right. \\
&{}\hspace{15pt}\left.- \mu_i \left(\frac{3}{8}\right)^k|\mathcal{Z}|^{2i}\right].
\end{IEEEeqnarray*}
We evaluate the dual at a point in its domain to obtain a lower bound to the optimal value of equation \eqref{opt:min_obj}. For $\bm{\nu}=\bm{0}$ and
\begin{numcases}{\mu_i=}
\lfloor ck \rfloor -i, & if $i\in \{1,2,\hdots,\lfloor ck \rfloor\}$ \nonumber\\
0, & otherwise,\nonumber
\end{numcases} where $c=\log_{|\mathcal{Z}|}\left(\frac{8}{3}\right)$, the value of the dual function is
\begin{IEEEeqnarray*}{rL}
&\mathcal{L}(\lambda)=\lambda -\left(\frac{3}{8}\right)^k\sum_{i=1}^{\lfloor ck \rfloor}(\lfloor ck\rfloor -i)|\mathcal{Z}|^{2i}\\
&{}-\Delta\sum_{i=1}^{\lfloor ck \rfloor}\exp\left(\frac{\lambda -\lfloor ck\rfloor-\Delta}{\Delta}\right)-\Delta\sum_{i > \lfloor ck \rfloor}\exp\left(\frac{\lambda -i-\Delta}{\Delta}\right).
\end{IEEEeqnarray*}
We can separately evaluate that
\begin{equation*}
\left(\frac{3}{8}\right)^k\sum_{i=1}^{\lfloor ck \rfloor}(\lfloor ck\rfloor -i)|\mathcal{Z}|^{2i}\leq\frac{|\mathcal{Z}|^2}{(|\mathcal{Z}|^2-1)^2} ~\text{for large}~k.
\end{equation*}
Using this and expanding the geometric series, we obtain that
\begin{IEEEeqnarray*}{rL}
\mathcal{L}(\lambda)\geq {}&\lambda - \frac{|\mathcal{Z}|^2}{(|\mathcal{Z}|^2-1)^2}-\Delta\sum_{i=1}^{\lfloor ck \rfloor}\exp\left(\frac{\lambda -\lfloor ck\rfloor-\Delta}{\Delta}\right)\\
&{}-\frac{\Delta}{e-e^{1-1/\Delta}}\exp\left(\frac{\lambda-\lfloor ck \rfloor}{\Delta}\right).
\end{IEEEeqnarray*}
If we choose $\lambda = ck/2$, we can see that the value of the dual function will be positive for large $k$. Hence, for any probability mass function that satisfies Lemma \ref{lemma:stopping_time}, we get that the value of $\E N-\frac{\Delta}{\log e} H(N) \geq 0$, i.e., $H(N)/\E N \leq \epsilon$. \hfill \IEEEQED

\section{Proof of Claim \ref{claim:entropy_support_set}}\label{app:support_arg}
%\begin{IEEEproof}
We first show that atleast one of the components of $\bm{q}_{m+1}$ is exactly equal to $c$. Pick any $q \in \mathcal{Q}_{m+1}$ and arrange its components in nonincreasing order so that $\bm{q}(m+1)$ is its smallest component. If $\bm{q}(m+1)>c$, then we can express $\bm{q}$ as a convex combination of two other elements of $\mathcal{Q}_{m+1}$ as follows. Note that $\bm{q}(1)\geq \bm{q}(m+1)\geq c$.
\begin{IEEEeqnarray*}{Rl}
\bm{q}=\frac{1}{2}\begin{bmatrix}
\bm{q}(1)+\bm{q}(m+1)-c\\
\bm{q}(2)\\
\vdots\\
\bm{q}(m)\\
c
\end{bmatrix}+\frac{1}{2}\begin{bmatrix}
\bm{q}(1)-\bm{q}(m+1)+c\\
\bm{q}(2)\\
\vdots \\
\bm{q}(m)\\
2\bm{q}(m+1)-c
\end{bmatrix}
\end{IEEEeqnarray*}
By concavity of entropy function we then have that this $\bm{q} \neq \bm{q}_{m+1}$. Thus $\bm{q}_{m+1}$ must have atleast one component that is equal to $c$, and WLOG let $\bm{q}_{m+1}=c$. Let $Q$ be a random variable on $[m+1]$ following the probability distribution $\bm{q}_{m+1}$. Let $I=\bm{1}_{\{Q=m+1\}}$ be the indicator function for the event $\{Q=m+1\}$. Then $\Pr(I=1)=c$ and $\Pr(I=0)=1-c$. Then we have that
\begin{IEEEeqnarray*}{Rl}
H(Q,I)=&H(Q)=H(I)+H(Q|I)\\
=&H(I)+\Pr(I=0)H(Q|I=0).
\end{IEEEeqnarray*}
Since $\Pr(Q|I=0)$ is a probability distribution over $[m]$ and $\Pr(Q=i|I=0)>\Pr(Q=i)$ for all $i \in [m]$, $\Pr(Q|I=0) \in \mathcal{Q}_m$. Hence we have that $H(Q|I=0)\geq H(\bm{q}_m)$ and using this in the previous equation, we get
\begin{IEEEeqnarray*}{C}
H(Q)\geq c\log \frac{1}{c}+(1-c)\log \frac{1}{1-c}+(1-c)H(\bm{q}_m).
\end{IEEEeqnarray*}
Since $mc<1$, we have that $-\log c>\log m$. But $\log m$ is the entropy of the uniform distribution over $[m]$ and hence $\log m \geq H(\bm{q}_m)$. Using this in the previous equation we get
\begin{equation*}
H(Q)\geq cH(\bm{q}_m)+(1-c)\log \frac{1}{1-c}+(1-c)H(\bm{q}_m)\geq H(q_{m}).
\end{equation*}
The same argument can be repeated to show that $H(\bm{q}_{m+1})\geq H(\bm{q}_{m}) \geq H(\bm{q}_{m-1})$ and so on. Hence, any probability mass function that satisfies a lower bound on the values of each of its probability masses necessarily has an equal or larger entropy if it is nonzero over a larger set. \hfill \IEEEQED
%\end{IEEEproof}
\section{Proof of Claim \ref{claim:convex_comb}}\label{app:convex_comb}
We show that any $\bm{p} \in $ conv($\mathcal{P}$) can be written as a convex combination of permuted $\bm{p}^\star$'s. Since $\bm{p} \in \text{conv}(\mathcal{P})$, we can write that $\bm{p} =\sum_i\mu_i\bm{p}_i$ where each $\bm{p}_i \in \mathcal{P}$, each $\mu_i \geq 0$ and $\sum_i \mu_i =1$. One can then see that if the claim is true for each $\bm{p}_i$ then it is also true for $\bm{p}$. Hence we focus on $\bm{p} \in \mathcal{P}$ and show that above claim holds for it. Without loss of generality, we can assume that the target vector $\bm{p}$ is arranged in non-increasing order, otherwise we permute its components so that the largest component is the first component and the successive components are in a non-increasing order. We can then reverse the permutation for every vector in its convex combination finally to get back our original vector.

Algorithm \ref{alg:convex_comb} returns a list of vectors, each of which can be expressed as a convex combination of permuted $\bm{p}^\star$'s. Using this list, we can find the convex combination of permuted $\bm{p}^\star$'s for any given $\bm{p} \in \mathcal{P}$ arranged in non-increasing order. The notation $\bm{p}'[i \leftrightarrow j]$ indicates the vector $\bm{p}'$ with its values at the $i$th and $j$th components interchanged, while all the other components remain the same.
\begin{algorithm}
\caption{Convex combination of permuted $\bm{p}^\star$'s for $\bm{p}$.}\label{alg:convex_comb}
\begin{algorithmic}[1]
\REQUIRE $\bm{p} \in$ conv($\mathcal{P}$) arranged in nonincreasing order, $\bm{p}^\star$.
\ENSURE A list $L$ of vectors.
\STATE Initialize $\bm{p}' \leftarrow \bm{p}^\star, L \leftarrow \phi$. $\bm{v}, \lambda$ are temporary variables.
\WHILE{$\bm{p}'-\bm{p}\neq\bm{0}$}
\STATE Find the smallest indices $i,j$ such that $\bm{p}'(i)>\bm{p}(i)$ and $\bm{p}'(j)<\bm{p}(j)$.
\STATE Evaluate
\begin{equation*}
\lambda :=\frac{\min \{\bm{p}'(i)-\bm{p}(i), \bm{p}(j)-\bm{p}'(j)\}}{\bm{p}'(i)-\bm{p}'(j)}.
\end{equation*}
\STATE Add the vector $\bm{v} := (1-\lambda)\bm{p}'+\lambda \bm{p}'[i\leftrightarrow j]$ to the list $L$.
\STATE Update $\bm{p}' \leftarrow \bm{v}$.
\ENDWHILE
\end{algorithmic}
\end{algorithm}

Intuitively, the algorithm finds the difference between $\bm{p}$ and $\bm{p}^\star$ and computes an intermediate vector $\bm{v}$ that can be written as a convex combination of permuted $\bm{p}^\star$'s. Following this, it finds the new difference between $\bm{v}$ and the target vector $\bm{p}$ and repeats the previous procedure. Finally, it stops when the intermediate vector equals $\bm{p}$. The correctness of the algorithm is ensured by the following claims.
\begin{claim} \label{claim:step5}
Let $m:=\min \{\bm{p}'(i)-\bm{p}(i), \bm{p}(j)-\bm{p}'(j)\}$. Then, at step 5 in algorithm \ref{alg:convex_comb},
\begin{IEEEeqnarray*}{C}
\bm{v}(i)=\bm{p}'(i)-m,\;\text{and}\; \bm{v}(j)=\bm{p}'(j)+m.
\end{IEEEeqnarray*} Also $\bm{v}(k)=\bm{p}'(k)$ for all $k \neq i,j$.
\end{claim}
\begin{IEEEproof}
Substituting $\lambda=m/(\bm{p}'(i)-\bm{p}'(j))$ gives the result.
\end{IEEEproof}
\begin{claim}\label{claim:step3}
At step 3 in algorithm \ref{alg:convex_comb}, $i<j$ and $\sum_{k=1}^{u}(\bm{p}'(k)-\bm{p}(k))\geq 0$ for all $u \geq j$.
\end{claim}
\begin{IEEEproof}
We prove this by induction on the iteration number of the WHILE loop. Suppose the indices $i,j$ found at the $t$th iteration be $i_t,j_t$ and the $\bm{v}$ evaluated at step 5 be denoted by $\bm{v}_t$. Then by Claims \ref{claim:t} and \ref{claim:step5}, we conclude that $i_1<j_1$ and $\sum_{k=1}^{u}(\bm{v}_1(k)-\bm{p}(k))\geq 0$ for $u\geq j_1$. As induction hypothesis, we assume that $i_t<j_t$ and $\sum_{k=1}^{u}(\bm{v}_t(k)-\bm{p}(k))\geq 0$ for $u \geq j_t$.

If at the $t$th iteration, $m=\bm{p}'(i_t)-\bm{p}(i_t)$, then by Claim \ref{claim:step5}, we get that $\bm{v}_t(i_t)-\bm{p}(i_t)=0$ and $\bm{v}_t(j_t)-\bm{p}(j_t)\leq 0$. Also $\bm{v}_t(k)-\bm{p}(k)=\bm{p}'(k)-\bm{p}(k)$ for all $k \neq i_t,j_t$. This implies that at the $(t+1)$th iteration, $j_{t+1}=j_t$. Furthermore, note that $\bm{v}_{t+1}$ and $\bm{v}_t$ differ only at the indices $i_{t+1}$ and $j_{t+1}$. For all $u\geq j_{t+1}=j_t$
\begin{equation*}
\sum_{k=1}^{u}(\bm{v}_{t+1}(k)-\bm{p}(k))=\sum_{k=1}^{j_t}(\bm{v}_t(k)-\bm{p}(k))\geq 0.
\end{equation*}
Since $\bm{v}_{t+1}(j_{t+1})=\bm{v}_t(j_t)$ and $\bm{v}_t(j_t)-\bm{p}(j_t)\leq 0$ it must be true that $i_{t+1}<j_{t+1}$.

On the other hand, if $m=\bm{p}(j_t)-\bm{p}'(j_t)$ at the $t$th iteration, then similarly $\bm{v}_t(j_t)-\bm{p}(j_t)=0, \bm{v}_t(i_t)-\bm{p}(i_t)\geq 0$ and $\bm{v}_t(k)-\bm{p}(k)=\bm{p}'(k)-\bm{p}(k)$ for all $k \neq i_t,j_t$. This implies that $i_{t+1}=i_t$ and $j_{t+1}>j_t$. Since $i_{t+1}<j_{t+1}$ and these are the only two indices affected at the $(t+1)$th iteration, we have that for all $u \geq j_{t+1}$
\begin{equation*}
\sum_{k=1}^{u}(\bm{v}_{t+1}(k)-\bm{p}(k))=\sum_{k=1}^{u}(\bm{v}_t(k)-\bm{p}(k))\geq 0
\end{equation*} as $u \geq j_{t+1}>j_t$. This completes the induction step and proves our claim.
\end{IEEEproof}
\begin{claim}
At step 4 of the algorithm, $\lambda \in [0,1]$.
\end{claim}\label{claim:step4}
\begin{IEEEproof}
By definition of the indices $i,j$, we have that $\bm{p}'(i)>\bm{p}(i)$ and $\bm{p}(j)>\bm{p}'(j)$. From Claim \ref{claim:step3}, we have that $i<j$. Also, by assumption, $\bm{p}$ is arranged in nonincreasing order. This implies that $\bm{p}(i)\geq \bm{p}(j)$. That gives the following string of inequalities
\begin{equation*}
\bm{p}'(i)>\bm{p}(i)\geq \bm{p}(j)>\bm{p}'(j).
\end{equation*}
This implies that $\lambda \geq 0$. In addition it also implies that
\begin{IEEEeqnarray*}{Rl}
\bm{p}(i)\geq \bm{p}'(j) \Leftrightarrow &\bm{p}'(i)-\bm{p}(i)\leq \bm{p}'(i)-\bm{p}'(j),\\
\bm{p}(j)\leq \bm{p}'(i) \Leftrightarrow &\bm{p}(j)-\bm{p}'(j)\leq \bm{p}'(i)-\bm{p}'(j).
\end{IEEEeqnarray*} These imply that $\lambda \leq 1$.
\end{IEEEproof}
The above claims conclude that $\bm{v}$ is a convex combination of vectors from conv($\mathcal{P}$). In the following claim we prove that $\bm{v}$ finally converges to $\bm{p}$ in a bounded number of steps.
\begin{claim}
The WHILE loop in algorithm \ref{alg:convex_comb} terminates after a bounded number of iterations.
\end{claim}
\begin{IEEEproof}
From calculations in Claim \ref{claim:step3} we concluded that at the end of the $t$th iteration, either $\bm{p}'(i_t)=\bm{p}(i_t)$ and/or $\bm{p}'(j_t)=\bm{p}(j_t)$ based on the value of $m$. Thus $\bm{p}'-\bm{p}$ will have a zero element at atleast one of $i_t$ or $j_t$. Also, none of the indices for which the difference $\bm{p}'-\bm{p}$ is already zero are affected by the algorithm as the inequality at step 3 is strict. Thus, the number of zero elements in the vector $\bm{p}'-\bm{p}$ increases by atleast one in every successive iteration. Since there are a finite number of components, it finally stops when the difference is the zero vector.
\end{IEEEproof}

\section{}\label{app:clumpy_entropy}
Note that expression for entropy in equation \eqref{eq:clumpy_entropy} can be expressed as follows
\begin{IEEEeqnarray}{Rl}
H(\bm{p}^\star)=&\frac{1}{M_{x,y}}\sum_{j=1}^{x+y+1}\lceil 2^{x+y-j}\rceil \left(\sum_{s=0}^{j-1}\right)\binom{x+y}{s}B(x,y,j) \IEEEeqnarraynumspace \label{eq:lower_bd_expr_A}
\end{IEEEeqnarray} where
\begin{equation*}
B(x,y,j)=(x+y)\log 3-\log \sum_{t=0}^{j-1}\binom{x+y}{t}.
\end{equation*} For a lower bound, we ignore the ceiling and simplify parts of the RHS of equation \eqref{eq:lower_bd_expr_A} as follows.
%We restrict our attention to parts of the RHS of equation \eqref{eq:lower_bd_expr_A} and simplify it as follows.
\begin{IEEEeqnarray*}{Rl}
&\sum_{j=1}^{x+y+1}2^{-j}\sum_{t=0}^{j-1}\binom{x+y}{t}\\
=&\frac{1}{2}\binom{x+y}{0}+\frac{1}{2^2}\left[\binom{x+y}{0}+\binom{x+y}{1}\right]+\hdots\\
&{}+\frac{1}{2^{x+y+1}}\left[ \binom{x+y}{0}+\hdots+\binom{x+y}{x+y}\right],\\
=&\binom{x+y}{0}\left[\frac{1}{2}+\frac{1}{2^2}+\hdots+\frac{1}{2^{x+y+1}}\right]\\
&{}+\:\binom{x+y}{1}\left[\frac{1}{2^2}+\hdots+\frac{1}{2^{x+y+1}}\right]+\hdots \\
&{}+\binom{x+y}{x+y}\frac{1}{2^{x+y+1}},\\
=&\binom{x+y}{0}\left[\frac{1}{2^0}-\frac{1}{2^{x+y+1}}\right]+\binom{x+y}{1}\left[\frac{1}{2^1}-\frac{1}{2^{x+y+1}}\right]\\
&{}+\hdots+\binom{x+y}{x+y}\left[\frac{1}{2^{x+y}}-\frac{1}{2^{x+y+1}}\right],\\
=&\binom{x+y}{0}\frac{1}{2^0}+\binom{x+y}{1}\frac{1}{2^1}+\hdots+\binom{x+y}{x+y}\frac{1}{2^{x+y}}\\
&{}-\left[\binom{x+y}{0}+\binom{x+y}{1}+\hdots+\binom{x+y}{x+y}\right]\frac{1}{2^{x+y+1}},\\
=&\left(\frac{3}{2}\right)^{x+y}-\frac{1}{2}.
\end{IEEEeqnarray*}
We also have that
\begin{IEEEeqnarray*}{Rl}
&\sum_{j=1}^{x+y+1}2^{-j}\left(\sum_{s=0}^{j-1}\binom{x+y}{s}\right)\log \sum_{t=0}^{j-1}\binom{x+y}{t}\\
=&\frac{1}{2}\binom{x+y}{0}\log\binom{x+y}{0}\\
&{}+\frac{1}{2^2}\left[\binom{x+y}{0}+\binom{x+y}{1}\right]\log\left[\binom{x+y}{0}+\binom{x+y}{1}\right]\\
&{}+\hdots +\frac{1}{2^{x+y+1}}\left[\sum_{s=0}^{x+y}\binom{x+y}{s}\right]\log \sum_{t=0}^{x+y}\binom{x+y}{t},\\
=&\binom{x+y}{0}\Bigg(\frac{1}{2}\log\binom{x+y}{0}+\frac{1}{2^2}\log\left[\binom{x+y}{0}+\binom{x+y}{1}\right]\\
&{}\hfill +\hdots+\frac{1}{2^{x+y+1}}\log \left[\binom{x+y}{0}+\hdots+\binom{x+y}{x+y}\right]\Bigg)\\
&{}+\binom{x+y}{1}\Bigg(\frac{1}{2^2}\log\left[ \binom{x+y}{0}+\binom{x+y}{1}\right]+\hdots\\
&{}\hfill +\frac{1}{2^{x+y+1}}\log\left[\binom{x+y}{0}+\hdots+\binom{x+y}{x+y}\right]\Bigg)\\
&{}+\hdots +\binom{x+y}{x+y}\frac{1}{2^{x+y+1}}\left[\binom{x+y}{0}+\hdots+\binom{x+y}{x+y}\right],\\
<&\binom{x+y}{0}\Bigg[\frac{1}{2}\log 2^{x+y}+\frac{1}{2^2}\log 2^{x+y}\\
&{}\hspace{120pt} +\hdots+\frac{1}{2^{x+y+1}}\log 2^{x+y}\Bigg]\\
&{}+\binom{x+y}{1}\left[\frac{1}{2^2}\log 2^{x+y}+\hdots+\frac{1}{2^{x+y+1}}\log 2^{x+y}\right]\\
&{}+\hdots+\binom{x+y}{x+y}\frac{1}{2^{x+y+1}}\log 2^{x+y},\\
=&\binom{x+y}{0}(x+y)\left[\frac{1}{2^0}-\frac{1}{2^{x+y+1}}\right]\\
&{}+\binom{x+y}{1}(x+y)\left[\frac{1}{2^1}-\frac{1}{2^{x+y+1}}\right] +\hdots\\
&{}+\binom{x+y}{x+y}(x+y)\left[\frac{1}{2^{x+y}}-\frac{1}{2^{x+y+1}}\right],\\
=&(x+y)\left[\left(\frac{3}{2}\right)^{x+y}-\frac{1}{2}\right].%
\end{IEEEeqnarray*}
%=\binom{x+y}{0}\frac{x+y}{2}\left[1+\frac{1}{2}+\hdots+\frac{1}{L_{x,y}}\right]\\
%\hfill +\binom{x+y}{1}\frac{x+y}{2^2}\left[1+\frac{1}{2}+\hdots+\frac{1}{2^{x+y-1}}\right]\\
%\hfill +\hdots+\binom{x+y}{x+y}\frac{x+y}{2^{x+y+1}}\\
Putting the parts together, we get that
\begin{IEEEeqnarray*}{Rl}
H(\bm{p}^\star)\geq &\frac{(x+y)L_{x,y}}{M_{x,y}}(-1+\log 3)\left[ \left( \frac{3}{2}\right)^{x+y}-\frac{1}{2}\right].
\end{IEEEeqnarray*}
\section{}\label{app:average_clumpy_entropy}
%Note that $\Pr(\bm{\Sigma}=\bm{\sigma})=M_{x,y}/8^k$ and the number of different $\bm{\sigma}$'s having the same number of 1's and 2's ($x$ and $y$ respectively) is $\frac{k!2^{k-x-y}}{x!y!(k-x-y)!}$. For our lower bound, we assume that the conditional probability mass function for all the $\bm{\sigma}$'s with equal number of 1's and 0's is the same. We use these in inequation \eqref{eq:lower_bd_expr_A} to obtain the following.
We use the bound obtained in Appendix \ref{app:clumpy_entropy} for the entropy of the clumpy distribution to obtain the following.
\begin{IEEEeqnarray*}{l}
H(\bm{Z}_1^N,\bm{Z}_2^N|\bm{\Sigma})=\sum_{\bm{\sigma}}\Pr(\bm{\Sigma}=\bm{\sigma})H(\bm{Z}_1^N,\bm{Z}_2^N|\bm{\Sigma}=\bm{\sigma})\\
\geq \sum_{\begin{matrix}
x,y=0\\
x+y \leq k
\end{matrix}}^{k}\Bigg(\frac{k!2^{k-x-y}}{x!y!(k-x-y)!}\left(\frac{1}{8}\right)^{k-x-y}\left(\frac{3}{8} \right)^{x+y}\\
\hspace{60pt} \times \frac{(x+y)(-1+\log 3)L_{x,y}}{M_{x,y}}\left[\left(\frac{3}{2}\right)^{x+y}-\frac{1}{2}\right]\Bigg),\\
\geq \sum_{\begin{matrix}
x,y=0\\
x+y \leq k
\end{matrix}}^{k}\frac{k!(\log 3-1)(x+y)}{x!y!(k-x-y)!4^k}\left[\left(\frac{3}{2}\right)^{x+y}-\frac{1}{2}\right],\\
=\frac{1}{4^k}\left[\sum_{\begin{matrix}
x,y=0\\
x+y \leq k
\end{matrix}}^{k}\frac{k!(x+y)(\log 3-1)}{x!y!(k-x-y)!}\left(\frac{3}{2}\right)^{x+y}\right.\\
\hfill \left. -\sum_{\begin{matrix}
x,y=0\\
x+y \leq k
\end{matrix}}^{k}\frac{k!(\log 3-1)(x+y)}{x!y!(k-x-y)!2}\right]\\
=\frac{\log 3-1}{4^k}\left[2.\frac{3}{2}.k.4^{k-1}-\frac{1}{2}.2.k3^{k-1} \right],\\
=\frac{\log 3-1}{4^k}\left[\frac{3k}{4}4^k-\frac{k}{3}3^k\right],\\
\implies \frac{H(\bm{Z}_1^N,\bm{Z}_2^N|\bm{\Sigma})}{k}\rightarrow 0.75(\log 3-1) ~\text{as}~ k \rightarrow \infty. %&\IEEEQEDhere
\end{IEEEeqnarray*}
\end{document}